\documentclass[11pt]{article}
\usepackage{epsfig,sint,macros,cite}
\usepackage{amssymb}
\begin{document}
\begin{titlepage}
\begin{flushright}
  DESY-99-023\\
  OUTP-99-07P
\end{flushright}

\vskip 0.5 cm
\begin{center}
  {\Large\bf 
  Hadron masses and matrix elements from  \\[0.5ex]
  the QCD Schr\"odinger functional\\[0.5ex] }
\end{center}
\vskip 0.5 cm
\vbox{
\centerline{
\epsfxsize=2.5 true cm
\epsfbox{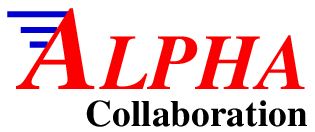}}
}
\vskip 0.5 cm
\begin{center}
{\large Marco Guagnelli$^{\scriptscriptstyle a}$, 
     Jochen Heitger$^{\scriptscriptstyle b}$,
     Rainer Sommer$^{\scriptscriptstyle b}$ and
     Hartmut Wittig$^{\scriptscriptstyle c,}$\footnote{PPARC Advanced Fellow}
\vskip 0.5cm
$^{\scriptstyle a}$
Dipartimento di Fisica, Universit\`a di Roma {\it Tor Vergata} \\
and INFN, Sezione di Roma II
\vskip 2.5ex
$^{\scriptstyle b}$
DESY-Zeuthen \\
Platanenallee 6, D-15738 Zeuthen, Germany
\vskip 2.5ex
$^{\scriptstyle c}$
Theoretical Physics, University of Oxford \\
1~Keble Road, Oxford OX1~3NP, UK
\vskip 1.0cm
{\bf Abstract}}
\vskip 0.7ex
\end{center}

We explain how masses and matrix elements can be computed in lattice QCD using
Schr\"odinger functional boundary conditions. Numerical results in the quenched 
approximation demonstrate that good precision can be achieved. 
For a statistical sample of the same size, our
hadron masses have a precision similar to
what is achieved with standard methods, but for the 
computation of matrix elements such as the pseudoscalar decay constant
the Schr\"odinger functional 
technique turns out to be much more efficient than the known alternatives.
\vfill

\begin{center}
March 1999
\end{center}

\eject

\vfill

\eject

\end{titlepage}

\section{Introduction}

In recent years, lattice QCD calculations in the quenched
approximation have reached a new
quality~\cite{reviews:leshouches,reviews:kenway98,lat98:ruedi}. The
renormalization of many local composite operators can be treated
non-perturbatively, and the leading discretization errors have been
removed. Consequently one would now like to perform continuum
extrapolations for various hadron masses and matrix elements in the
improved theory
\cite{impr:qcdsf,impr:roma2_2,impr:roma2_3,impr:scri,impr:roma1_spec,impr:ukqcd_prep}
and compare to recent results obtained with more standard
techniques~\cite{lat98:ruedi}. In particular, the problem of
a determination of the light quark masses can be addressed with confidence,
since the complete renormalization is known
non-perturbatively~\cite{mbar:pap1}.

However, continuum extrapolations require masses and matrix elements
for finite values of the lattice spacing which are sufficiently
accurate~\cite{lat98:ruedi}. Efficient methods are needed to obtain
such precise results. Standard methods use correlation functions in
a sufficiently large box with periodic boundary
conditions~\cite{rev:rajan}. Usually one seeks to enhance the
dominance of the low lying hadrons by using tuned hadron wave
functions
\cite{smear:cube,smear:wall,smear:DeGrand1,smear:gauss,smear:Degrand2,fbstat:FNAL,smear:wupp,fbstat:npb94,UKQCD:jacobi,UKQCD:fuzzing},
possibly combined with variational techniques
\cite{var:michael85,phaseshifts:LW,fbstat:FNAL}. 

In this paper we investigate an alternative to these standard
methods. Again we choose a sufficiently large box but impose
Dirichlet boundary conditions in time, as they are used to formulate
the QCD {\SF}. We shall demonstrate that correlation functions in
the \SF are dominated by hadron intermediate states at large
Euclidean time. Moreover, it is shown that a time extent of 3~fm for
the box is sufficient to extract masses and matrix elements. An
advantage compared to the standard methods is that the
pre-asymptotic decay of \SF correlation functions is very slow,
which means that a large signal remains at large separations.

In the following, we briefly discuss the foundation of the method
(Section \ref{s_corr}) and test its applicability in practice in the
quenched approximation (Section \ref{s_extr}). We also attempt to
quantify the efficiency compared to more standard calculations
(Section \ref{s_eff}). Finally we discuss open questions as well as
possible further improvements.

\section{Correlation functions at large time separations \label{s_corr}}

We now derive explicit expressions for the representation
of Schr\"odinger functional correlation functions in terms of
intermediate physical states. Throughout this section we assume that
the lattice \SF\ is defined using the standard Wilson action as in
ref.~\cite{SF:stefan1}. In this situation the relations presented here
hold exactly. 
If one considers the $\Oa$ improved theory, as we will do later, we
cannot derive the equations given in this section directly from the
transfer matrix. However, universality implies that the renormalized
correlation functions of the improved theory and the unimproved
theory agree in the continuum limit. 
Since the correlation functions considered below are renormalized
multiplicatively, their time dependence (bare or renormalized) is
given by the expressions derived for the Wilson theory up to lattice
spacing effects. For the $\Oa$ improved theory this means that all
relations derived in this section are valid for physical distances
large compared to the lattice spacing (up to corrections of
order~$a^2$).

The correlation functions considered here have been introduced
before~\cite{impr:lett,impr:pap1,impr:pap5}. In those references the
emphasis was largely on the perturbative regime, which means
choosing small extensions of the space-time volume. By contrast, in
this work we are interested in the correlation functions for
intermediate to large volumes, i.e. extensions which are
significantly larger than typical QCD scales. Provided that the pion
mass is not too small, such typical QCD scales 
are of order $1~\fm$.

The QCD \SF is defined as the QCD partition function in a cylindrical
geometry, i.e. periodic boundary conditions with periodicity length
$L$ in three of the four Euclidean dimensions, and Dirichlet boundary
conditions in time at the hypersurfaces $x_0=0$ and $x_0=T$. Its
quantum mechanical interpretation has first been discussed for the pure
gauge theory \cite{SF:LNWW} and subsequently for the theory with
quarks~\cite{SF:stefan1}. In these references it was shown that the
\SF partition function can be written as
\bes
 {\pf} = \langle {\rm f}| \rme^{-T \ham } \projector |{\rm i}\rangle
 \label{e_Z}\, ,
\ees
with states $|{\rm i}\rangle$ and $|{\rm f}\rangle$ which are given in
terms of the boundary values specified at $x_0=0$ and $x_0=T$,
respectively. In the above equation $\ham$ denotes the Hamilton
operator of QCD formulated on a torus of volume $L^3$. More precisely,
in the lattice theory it is proportional to the (negative) logarithm
of the transfer matrix.\footnote{
%
%
For the unimproved Wilson action, $\ham$ is known to be 
hermitian~\cite{Luscher:TM}.}

Of course, the same operator describes the correlation functions
when the Dirichlet boundary conditions are replaced by periodic
boundary conditions in time. The projector $\projector$ projects
onto the gauge invariant subspace of the Hilbert
space~\cite{SF:LNWW,SF:stefan1}; only gauge invariant intermediate
states are physical and can contribute.

For our present investigation, we have considered only the case of
homogeneous boundary conditions, where the spatial components of the
gauge potentials are set to zero at the boundaries and also the
fermion boundary fields are taken to vanish. In this case we have
$|{\rm i}\rangle = |{\rm f}\rangle = \initial$ and this state
carries the quantum numbers of the vacuum. Other choices for the
boundary conditions may be of interest as well and can be treated
similarly.

As an example we will discuss two specific correlations, which
allow for a calculation of the pion mass and decay constant.  The
generalization to other channels and other matrix elements is
straightforward.

We start from the dimensionless fields
\bes
 \op{} = {{a^{6}}\over{L^3}} \sum_{\vecy, \vecz}
           \zetabar_{\rm u}(\vecy)\gamma_5\zeta_{\rm d}(\vecz) 
        \,, \quad
 \op{}' = {{a^{6}}\over{L^3}} \sum_{\vecy, \vecz}
           \zetabarprime_{\rm d}(\vecy)\gamma_5
                            \zetaprime_{\rm u}(\vecz) 
			    \label{e_op}
\ees
and a local composite (gauge invariant) field $X(x)$ (which will have mass
dimension three in our applications) to define
the gauge invariant correlation functions
\bes
  \fx(x_0)  & = & - {L^3 \over 2}\langle
                 X(x) \, \op{}
         \rangle \label{e_fa} \,,
         \\
  f_1 & = & -
         {1 \over 2}\langle \op{}' \, \op{} \rangle \label{e_f1}\,,
\ees
where the average denotes the usual path integral average and ${\rm
u,d}$ are flavour indices. The ``boundary quark fields'', $\zeta,
\ldots, \zetabarprime$, have been discussed in~\cite{impr:pap1}.
In the lattice theory, $\zeta$ is given explicitly in terms of the
gauge fields connecting hypersurfaces $x_0=0$ and $x_0=a$ and the
quark fields on the hypersurface $x_0=a$. Analogous properties hold
for the other boundary quark fields. Our choices for $X$ are
$X=A_0$ (defining, through \eq{e_fa}, the correlation function $\fa$)
and $X=P$ (which gives $\fp$), where
\bes
 A_0(x) &=& \psibar_{\rm d}(x)\gamma_0 \gamma_5\psi_{\rm u}(x)\,, \\
 P(x) &=&   \psibar_{\rm d}(x)         \gamma_5\psi_{\rm u}(x)\,. 
\label{e_pseudodef}
\ees
The correlation functions $\fx$
have the quantum mechanical representation
\bes
  \fx(x_0) = \pf^{-1}\,{L^3 \over 2}\,\initialt \rme^{-(T-x_0)\ham} 
                   \projector\opX  \rme^{-x_0 \ham }
                  \projector \initialpi \,, \;
		  a \leq x_0 \leq T-a \,,\label{e_fx_qu}
\ees
where $\opX$ is the corresponding operator in the Schr\"odinger
picture, and the state $\initialpi$ has the quantum numbers of the
$\pi^+$ with momentum zero. To conclude that \eq{e_fx_qu} holds, one
only requires that the combination of fields
$\op{}$, \eq{e_op}, has support for $x_0 \leq a$ and that it carries the
quantum numbers of a $\pi^+$. The former is guaranteed by the very
construction of the boundary fields $\zeta,\zetabar$~\cite{impr:pap1}. 
Furthermore we
have
\bes
  f_{1} = \pf^{-1}\,\frac12\,\initialpit \rme^{-T \ham } \projector
                   \initialpi  \label{e_f1_qu} \, .
\ees
It is now apparent that (for large separations $x_0$ and $T-x_0$) the
mass of the pion and its decay constant can be extracted. To see this
explicitly we insert a complete set of eigenstates of the Hamiltonian,
\bes
 &&|n,q\rangle\,,   \quad n=0,1,\ldots \;, \\
 && \ham \, |n,q\rangle = E^{(q)}_{n} |n,q\rangle \,,
\ees
with normalization
$
\langle n',q'|n,q\rangle = 
   \delta_{n,n'} \, \delta_{q,q'}
$.
Here the energy levels in the sector of the Hilbert space with
internal quantum numbers $q$ are enumerated by $n$.  Only quantum
numbers $q=\pi$, a shorthand for $(J,P,C,I,I_3) = (0,-,+,1,1)$, and
$q=0$, which denotes vacuum quantum numbers are considered in the
following. We do not indicate the momentum of the states
$|n,q\rangle$, since both $\initial$ and $\initialpi$ are invariant
under spatial translations and only states $|n,q\rangle$ with
vanishing (spatial) momentum contribute.

The representations given so far hold for arbitrary $L$, $T$ and
$x_0$. In this paper, we shall be interested in the special case of
the asymptotic behavior of $\fa(x_0),f_1$ for large values of both
$x_0$ and $T-x_0$, while $L$ remains unspecified at this stage. We
include the first non-leading corrections but neglect any
contributions which are suppressed by terms of order
\bes
 \exp(-T E^{(0)}_{1}), \; 
                 \exp(-x_0 E^{(\pi)}_{1} - (T-x_0) E^{(0)}_{1}), 
 \;
 \exp(-(T-x_0) E^{(0)}_{2}) \; {\rm and}
 \;
  \exp(-x_0 E^{(\pi)}_{2}) \,,  \nonumber
\ees
compared to the leading terms in the correlation functions.
In this approximation we obtain
\bes
  \fx(x_0) &\approx& 
     \frac{L^3}{2}  \, \rho \, \langle 0,0| \opX | 0,\pi\rangle   
                       \, \rme^{-x_0 m_{\pi} } \times
      \left\{ 1 + \etax^{\pi}\rme^{-x_0 \Delta } + 
                  \etax^{0} \rme^{-(T-x_0) m_{\rm G} } 
      \right\} , 
      \label{e_fa_asympt}\\
  f_{1} &\approx& \frac{1}{2}\,\rho^2 \, \rme^{-T m_{\pi} } 
                   \label{e_f1_asympt} \,, 
\ees
where we have introduced the ratios 
\bes                   
   \rho &=&  {\langle 0,\pi \initialpi 
             \over 
             \langle 0,0 \initial } ,
                       \label{e_rho}\\
 \etax^{\pi} &=& {\langle 0,0| \opX | 1,\pi\rangle \langle 1,\pi \initialpi
                   \over 
                 \langle 0,0| \opX | 0,\pi\rangle \langle 0,\pi \initialpi}
                   \,, \\
 \etax^{0} &=& { \initialt 1,0 \rangle \langle 1,0| \opX | 0,\pi\rangle 
                  \over
                 \initialt 0,0 \rangle \langle 0,0| \opX | 0,\pi\rangle }
            \,.              
\ees
The energy difference $m_{\rm G} = E^{(0)}_{1}-E^{(0)}_{0}$ is the
mass of the $0^{++}$ glueball and $\Delta=E^{(\pi)}_{1}-E^{(\pi)}_{0}$
is an abbreviation for the gap in the pion channel. As indicated
above, we have dropped contributions of higher excited states which
decay even faster as $x_0$ and $T-x_0$ become large.

Considering the special case of $\fa$, we find that it is
proportional to the matrix element $\langle 0,0| \opA |0,\pi\rangle$,
which is related to the pion decay constant $\Fpi$ through
\bes
    \za \langle 0,0| \opA | 0,\pi\rangle  = 
    \Fpi m_{\pi} (2m_{\pi}L^3)^{-1/2} \,.
\ees
Here, $\za$ is the renormalization constant of the isovector axial
current, and the factor $(2m_{\pi}L^3)^{-1/2}$ takes account of the
conventional normalization of one-particle states (in our convention
the experimental value of the pion decay constant is 132~MeV).

\Eq{e_fa_asympt} is used to determine $m_{\pi}$, while the pion
decay constant, $\Fpi$, may be conveniently extracted from the ratio 
\bes
 \za \, \fa(x_0) / \sqrt{f_{1}} &\approx& 
   \frac12 \Fpi \, (m_\pi  \,L^{3})^{1/2}
             \rme^{-(x_0-T/2) m_{\pi} } \nonumber \\
            && \times
      \left\{1 + \etaa^{\pi}\rme^{-x_0 \Delta } + 
                  \etaa^{0} \rme^{- (T-x_0) m_{\rm G}} 
      \right\} \, .
      \label{e_fpi}
\ees
The above formulas show explicitly how masses and matrix elements
can be obtained from \SF correlation functions. Before we describe
the numerical tests of the practicability, we wish to point out some
properties of the present method and list the differences to
conventional approaches.

\begin{itemize}
\item
     Equations (\ref{e_fa_asympt},\ref{e_f1_asympt}) can be expected
     to be rather accurate when all time separations are larger than
     typical hadronic length scales (say, $1\,\fm$, provided $m_\pi$
     is not too small). \SF correlation functions decay slowly for
     small $x_0$, leaving a large and precise signal at separations
     of $1-2\,\fm$.  This is easily seen by applying asymptotic freedom 
     and a
     simple dimensional analysis. \\ As a comparison, consider
     correlation functions of standard local composite fields such
     as $\sum_{\vecx} \langle \Phi^\dagger(x) \Phi(0) \rangle$,
     with $\Phi(x)=\psibar_{\rm u}(x) \Gamma \psi_{\rm d}(x)$,
     $\Phi^\dagger(x)=\psibar_{\rm d}(x)\gamma_0 \Gamma
     \gamma_0\psi_{\rm u}(x)$.  At distances $x_0 \approx 1\,\fm$,
     such correlation functions are typically very small which
     usually means low statistical precision. The reason why they
     are small is because they decay like $(x_0)^{-3}$ for short
     time separations, as may be inferred by the same arguments used
     above. This qualitative difference arises from the fact that in
     the \SF a dimensionless (non-local) field, $\int
     \rmd{\vecy}\,\,\zetabar_{\rm u}(\vecy)\gamma_5 \zeta_{\rm
     d}({\bf 0})$, is used to create hadronic states at the
     boundary.
\item     
     The ratio $\rho$, \eq{e_rho} is divergent, since the state
     $\initialpi$ involves the bare boundary quark fields.  However,
     in the final quantities of interest $\rho$ is cancelled
     explicitly (see \eq{e_fpi}).
\item
     In particular, the combination $\za \, f_{\rm A}(x_0) /
     \sqrt{f_{1}}$ has a continuum limit for all values of
     $x_0$. One may therefore choose some (not too large) value of
     the lattice spacing to determine the time separations $x_0$ and
     $T-x_0$ where the contamination due to excited states is small.
     The same separations (in physical units) can then be used for
     other values of the lattice spacing. This holds also for the
     ``local masses'' which are commonly used to extract hadron
     masses.  When one applies smearing or fuzzing the validity of
     such a statement is not immediately evident.
\item
     Spatial translation invariance is used fully and reduces 
     statistical errors.     
\item   
     The present approach is similar to using ``wall sources'' to
     create hadronic states \cite{smear:wall}, but here
     we keep gauge invariance in all stages of the formulation!
\end{itemize}

\section{Extraction of masses and decay constants \label{s_extr}}

In this section we demonstrate the practicability of the method in the
case of quenched lattice QCD.  

\subsection{Computational details}

We work in $\Oa$ improved QCD as detailed in~\cite{impr:pap1},
using the non-perturbative estimates of the improvement coefficients
$\csw$ and $\ca$ reported in ref.~\cite{impr:pap3}.

The full $\Oa$ improvement of the \SF correlation functions requires
also $\Oa$ counterterms (with coefficients $\ct,\cttil$) at the
boundary \cite{impr:pap1}. These do not affect the lattice spacing
dependence of hadron masses and matrix elements, and therefore their
coefficients are not very important in the present context. The only
place where they do play a role is the size of the correction terms in
\eq{e_fa_asympt}, since the lattice spacing errors in the amplitude
ratios
$\etax^{\pi},\etax^0$ are $\Oasq$ when $\ct,\cttil$ are chosen
appropriately. Otherwise cutoff effects of order~$a$ remain in 
these
excited state contributions. We used one-loop estimates for
$\ct,\cttil$ \cite{alpha:SU3,impr:pap2}. For these values the $\Oa$ effects
are expected to be quite small \cite{mbar:pap1,impr:scaletest}.

In the pseudoscalar channel we consider the correlation functions
$\fp(x_0)$ and $\faimpr(x_0)$, where the latter is obtained from
\eq{e_fa} by inserting the time component of the improved axial
current~\cite{impr:lett,impr:pap1}, viz.
\bes
 (A_{\rm I})_\mu(x) = A_\mu(x) +a\ca\frac{1}{2}\left(\partial^*_\mu
                                          +\partial_\mu\right) P(x).
\label{e_axialimpdef}
\ees
The correlation function in the vector channel is defined by
\bes
  \kvimpr(x_0) & = & -a^6\sum_{\vecy,\vecz}\frac{1}{6}\sum_k\langle
         (\vimpr)_k(x) \;
         \zetabar_{\rm u}(\vecy)\gamma_k\zeta_{\rm d}(\vecz)
         \rangle \label{e_kv} \,,
\ees
where
\bes
         (\vimpr)_\mu(x) & = & \psibar_{\rm d}(x)\gamma_\mu\psi_{\rm u}(x) 
         + a \cv \frac{i}{2} (\partial_\nu+\partial_\nu^*)
         [\psibar_{\rm d}(x)\sigma_{\mu\nu}\psi_{\rm u}(x)]\,
\ees
is the $\Oa$ improved vector current. For the improvement coefficient
$\cv$ we have used the values from its non-perturbative determination
reported in~\cite{lat97:marco}.

We have chosen the parameters of our simulations such that $L\approx
1.5\,\fm$ and $T\approx 3\,\fm$. Our calculations have been
performed for four different values of the lattice spacing, but for
the purpose of demonstrating the practicability of the method we
restrict ourselves to results obtained at $\beta=6.0$ and
$\beta=6.2$. These couplings correspond to lattice spacings
$a=0.093\,\fm$ and $a=0.068\,\fm$, when $r_0=0.5\,\fm$ is used to
set the scale \cite{pot:r0,pot:r0_SU3}. For these parameters a
direct comparison with results obtained using conventional
methods~\cite{impr:qcdsf,impr:roma2_2,impr:roma1_spec,impr:ukqcd_prep}
can be made.

\begin{figure}[tb]
\hspace{0cm}
\vspace{-4.2cm}

\centerline{
\psfig{file=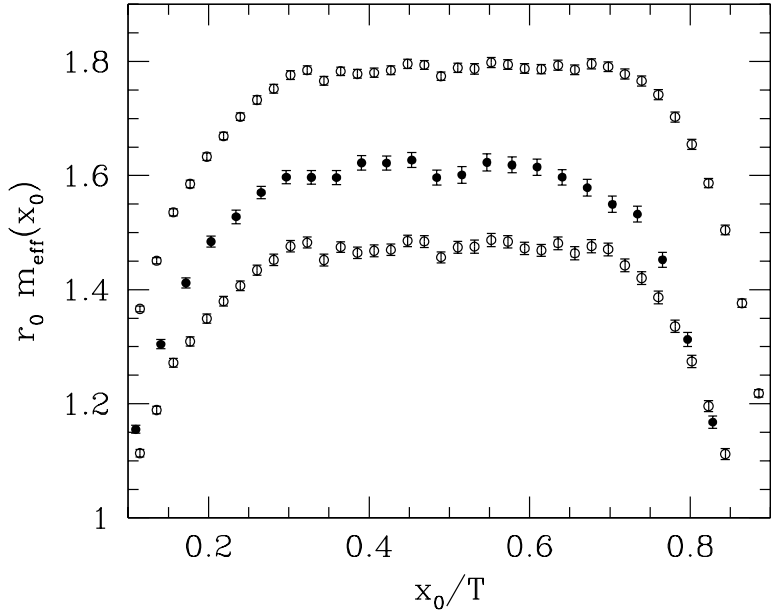,width=14cm}
}
\vspace{-1.2cm}
\caption{\footnotesize
The effective mass of the correlation function 
$\faimpr$, as defined in \eq{e_meff}.
The three sets of data points correspond to 
$(\beta, \kappa)=(6.2,0.1349),(6.0,0.1342),(6.2,0.1352)$ from top to bottom.
In all cases the time extent is $T\approx 3\,\fm$.
\label{f_meff_pi}}
\end{figure}
%
The numerical computations of \SF correlation functions have been
explained earlier \cite{impr:pap3}, and further details can be found
in that reference. Here we only mention that in addition to the
previously used even/odd preconditioned version of the BiCGStab
solver, we have also implemented SSOR-preconditioning
\cite{SSOR:wupp,SSOR:impr}. The latter reduced the number of BiCGStab
iterations needed to solve the Dirac equation by more than a
factor~2. This turned into a gain in CPU-time of a factor of around $1.5$
in our implementation on the APE-100 machines.

Our statistical samples consist of 1000 ``measurements'' of \SF
correlation functions at $\beta=6.0$ and 800 measurements at
$\beta=6.2$. All statistical errors were computed using the jackknife
method.

\subsection{Plateaux}

Let us first get a rough impression of the range of $x_0$, where the
leading term in \eq{e_fa_asympt} dominates. To this end we follow the
tradition of looking for a plateau in the effective mass, 
\bes
m_{\rm eff}(x_0+\frac{a}{2}) = {1 \over a}
\ln\left(f(x_0)/f(x_0+a)\right)\,.
\label{e_meff}
\ees
Here, $f$ denotes any of the two-point correlation functions defined
above. In the pseudoscalar channel we use $f=\faimpr$, and the
resulting effective masses are shown in \fig{f_meff_pi}. There is
good evidence for plateaux starting at $x_0 \approx 1\,\fm$ and
extending to approximately $T-x_0 \approx 1\,\fm$. As expected, the
location of the plateaux is approximately independent of the lattice
spacing.

\begin{figure}[tb]
\hspace{0cm}
\vspace{-4.2cm}

\centerline{
\psfig{file=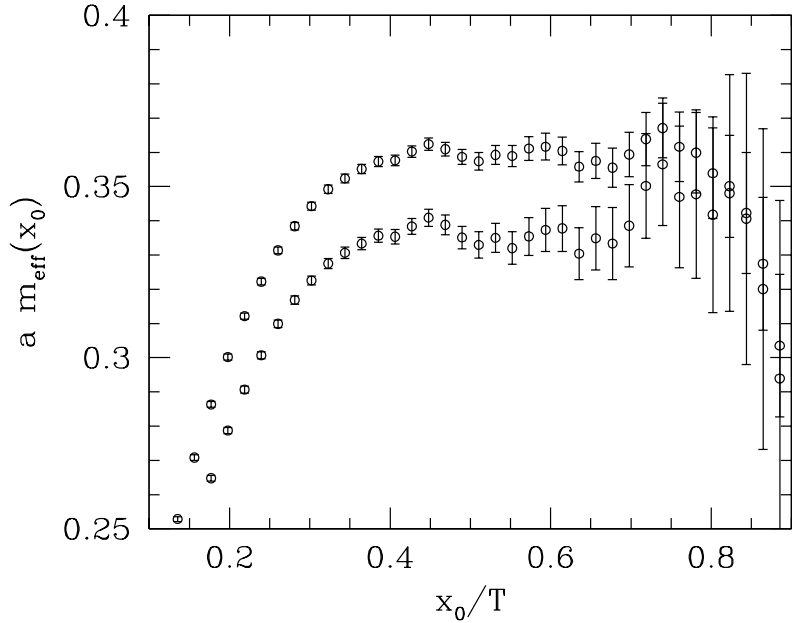,width=14cm}
}
\vspace{-1.2cm}
\caption{\footnotesize
The effective mass of the correlation function $\kvimpr$ for 
$a\approx 0.07\,\fm$. 
\label{f_meff_rho}}
\end{figure}

Turning to the vector channel, we use \eq{e_meff} with $f=\kvimpr$.
In this channel the effective mass, shown in \fig{f_meff_rho}, turns
into a plateau only at $x_0\approx 1.5\,\fm$. Furthermore, statistical
errors grow more rapidly as $x_0$ becomes large. By comparing
\fig{f_meff_pi} and \fig{f_meff_rho} one may anticipate that it is
more difficult to obtain a reliable determination of the vector meson
mass than it is to compute $m_{\pi}$. However, this is not much
different when standard methods are employed.

\subsection{Excited state corrections -- averaging intervals}

Let us now discuss the extraction of masses and matrix elements in
some detail. A standard method is to perform fits to the leading
term in \eq{e_fa_asympt}. Alternatively one can average the
effective mass over the plateau region. Both of these procedures
reduce the error in the mass compared to just taking one point of
the effective mass in the plateau. One must then ensure that the
final statistical error is not accompanied by a noticeable
systematic error due to a small contamination by excited states. In
other words, one needs a rough estimate of the size of the
contribution of excited states in the region of $x_0$ considered.

In order to arrive at such an estimate, one requires some
information about the values of $m_{\rm G}$ and the gap $\Delta$ in
\eq{e_fa_asympt} and the analogous equation for the vector channel.
Glueball masses from the literature \cite{reviews:glueballs}, combined
with $r_0/a$ \cite{pot:r0_SU3}, give
\bes
 m_{\rm G} r_0 \approx 4.3\,.
\ees 
In addition, we have obtained a rough estimate for $\Delta$ in the
pseudoscalar channel by analyzing correlation functions with
periodic boundary conditions in time, which were made available to
us by the {\it Tor Vergata} group \cite{impr:roma2_2}. Since this
analysis is not of immediate interest for our discussion of \SF
correlation functions we relegate the details to the appendix.

The typical uncertainties associated with the gaps and the scalar
glueball mass are of order $10\%$ for the range of quark masses considered
here. The gaps and $m_{\rm G}$ can now be used to take a closer look
at the effective masses and obtain estimates for the excited state
contributions and the range of $x_0$ where these are small compared to
the statistical errors. We discuss this separately for the two
channels.

\subsection{The pseudoscalar channel}

\begin{figure}[tb]
\hspace{0cm}
\vspace{-5.2cm}

\centerline{
\psfig{file=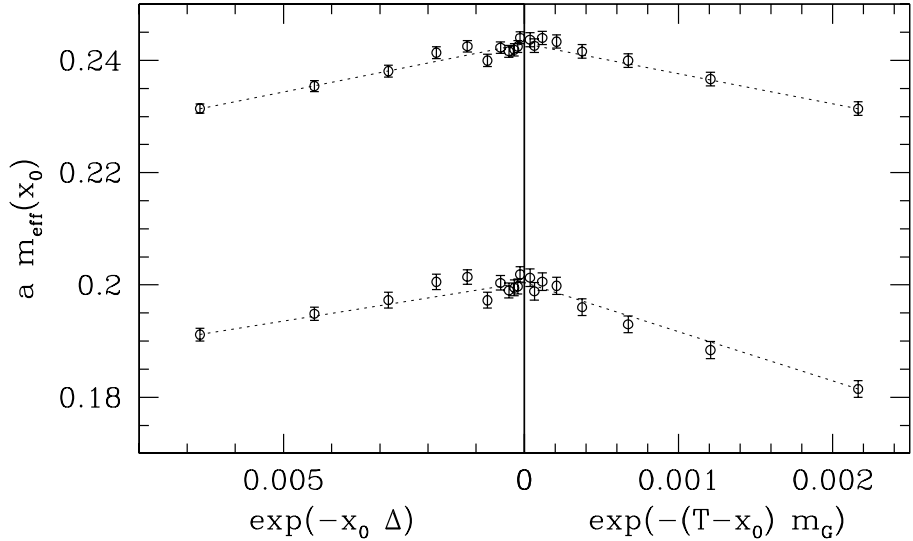,width=14cm}
}
\vspace{-1.2cm}
\caption{\footnotesize
The effective mass of the correlation function $\faimpr$ against the
two leading corrections. Small values of $x_0$ are omitted in the
right half of the graph and large values of $x_0$ in the left half.
Data are for a lattice spacing of $a \approx 0.07\,\fm$; the dashed
lines indicate the agreement with linear behaviour.
\label{f_m_eff_pi2}}
\end{figure}

The analysis described in the appendix yields
\bes
   r_0 \Delta  \approx 3.2\,,
\ees
which agrees with an estimate of the same quantity using data by
UKQCD~\cite{impr:ukqcd_prep}. Using \eq{e_fa_asympt},
the time dependence of the effective mass is given by
\bes
  \m_{\rm eff}(x_0) &\approx& \mpi\, 
      \left\{ 1 
        + \frac{2\sinh(a\Delta/2)}{a\mpi} \etaa^{\pi}\rme^{-x_0 \Delta } 
        - \frac{2\sinh(a m_{\rm G}/2)}{a\mpi} \etaa^{0} \rme^{-(T-x_0) m_{\rm G} } 
      \right\} \nonumber \\
      &\approx& \mpi\, 
      \left\{ 1 + \frac{\Delta}{\mpi} \etaa^{\pi}\rme^{-x_0 \Delta } 
                - \frac{m_{\rm G}}{\mpi} \etaa^{0} \rme^{-(T-x_0) m_{\rm G} }
      \right\}, \label{e_meff_asympt}
\ees
where the second line is valid close to the continuum limit ($a
\Delta \ll 1$, $a m_{\rm G}\ll 1$). In order to check whether this
time dependence is reproduced by the data, we plot the effective
masses directly against the expected form of the asymptotic
correction terms, $\rme^{-x_0 \Delta }$ and $\rme^{-(T-x_0) 
m_{\rm G}}$ in \fig{f_m_eff_pi2}. 

\begin{table}[b]
\centering
\begin{tabular}{cccc}
\hline \\[-1.0ex]
channel & $\eps$ & $t_{\rm min} / r_0$ & $(T-t_{\rm max}) / r_0$ 
\\[1.0ex]
\hline \\[-1.0ex]
  $\faimpr$& 0.2\% & 2.6 & 2.3 \\
 $\faimpr/\fp$, $\fp$ & 0.1\% & 2.8 & 2.5 \\
 $\kvimpr$ & 0.2\% & 3.0 & 2.2 \\[1.0ex]
\hline
\end{tabular}
\caption[TAB_zptree]{\footnotesize
Ranges of $x_0$, where (relative) excited state contribution are
smaller than $\eps$. 
\label{t_fitranges}
}
\end{table}

One observes that the data fall approximately onto straight
lines. It is important to bear in mind that $m_{\rm G}$, $\Delta$ on
the one hand and $m_{\rm eff}(x_0)$ on the other, have been obtained
independently from correlation functions computed for different
boundary conditions where excited state corrections have different
amplitudes. The apparent compatibility of the data with the expected
form of the correction terms is therefore non-trivial, and we
conclude that we have a semi-quantitative understanding of the
excited state corrections.

It is now easy to deduce values for $t_{\rm min}$ and $T-t_{\rm
max}$ such that for $t_{\rm min}\leq x_0 \leq t_{\rm max}$ these
corrections are below a certain margin, $\eps$, which is allowed as
a systematic uncertainty. We list $t_{\rm min}$ and $T-t_{\rm max}$
together with the chosen values for $\eps$ in
\tab{t_fitranges}. Reliable estimates for $\mpi$ are then obtained
by averaging $m_{\rm eff}$ for $t_{\rm min}\leq x_0 \leq t_{\rm max}$,
and representative results are collected in \tab{t_pion}.

A widely used method to extract hadron masses is to fit the
correlation functions, using the expected asymptotic behaviour as an
ansatz. In order to check the consistency of results obtained by
averaging $m_{\rm eff}$ for $t_{\rm min} \leq x_0 \leq t_{\rm max}$,
we have also performed single exponential fits to $\faimpr(x_0)$
over the same interval. A comparison of the two methods shows that
the estimates for $m_\pi$ are entirely consistent. Furthermore, the
quality of the fits is very good, with typical values of the
correlated $\chi^2/n_{\rm df}$ in the range $0.6-0.9$. Moreover, the
stability of the fits under variations of the fitting interval has
been used as an additional check that our values of $t_{\rm
min},\,t_{\rm max}$ were chosen appropriately. In \tab{t_pion} we
compare our results for $am_\pi$ to those obtained using
conventional techniques and find good agreement.
  
\begin{table}[tb]
\centering
\begin{tabular}{ccr@{.}lr@{.}lr@{.}lr@{.}l}
\hline \\[-1.0ex]
$\beta$ & $\kappa$  & \multicolumn{2}{c}{$a m_\pi$} &
\multicolumn{2}{c}{ref.\cite{impr:qcdsf}} & 
\multicolumn{2}{c}{ref.\cite{impr:roma2_2}} &
\multicolumn{2}{c}{ref.\cite{impr:roma1_spec}}\\[1.0ex] 
\hline \\[-1.0ex]
 6.0 & 0.1338 & 0&3529(11) & \blank & \blank     & \blank    \\
     & 0.1342 & 0&3001(12) & 0&2988(17) & \blank     & \blank    
              \\[1.0ex]
\hline \\[-1.0ex]
 6.2 & 0.1349 & 0&2430(6)  & 0&2444(9)  & \blank     & 0&2440(21)
              \\
     & 0.1352 & 0&2004(6)  & 0&2016(11) & 0&2007(40) & 0&2007(26)
              \\[1.0ex]
\hline
\end{tabular}

\vspace{1.0cm}

\begin{tabular}{ccr@{.}lr@{.}lr@{.}lr@{.}l}
\hline \\[-1.0ex]
$\beta$ & $\kappa$  & 
\multicolumn{2}{c}{$m_\pi\Fpi^{\rm bare} \over \Gpi^{\rm bare}$} &
\multicolumn{2}{c}{~~~~~$a\Fpi^{\rm bare}$} &
\multicolumn{2}{c}{ref.\cite{impr:qcdsf}} & 
\multicolumn{2}{c}{ref.\cite{impr:roma1_spec}}\\[1.0ex] 
\hline \\[-1.0ex]
 6.0 & 0.1338 & 0&2125(7) &   ~~~~~0&0943(4) & \blank    & \blank     \\
     & 0.1342 & 0&1794(8) &   ~~~~~0&0905(4) & 0&0907(8) & \blank     \\[1.0ex]
\hline \\[-1.0ex]
 6.2 & 0.1349 & 0&2165(5) &   ~~~~~0&0721(3) & 0&0727(9) & 0&0740(35) \\
     & 0.1352 & 0&1769(5) &   ~~~~~0&0687(3) & 0&0690(30) & 0&0706(46) \\[1.0ex]
\hline
\end{tabular}

\caption[]{\footnotesize
Results in the pseudoscalar channel compared to
values from the literature
\label{t_pion}
}
\end{table}

\begin{figure}[tb]
\hspace{0cm}
\vspace{-5.2cm}

\centerline{
\psfig{file=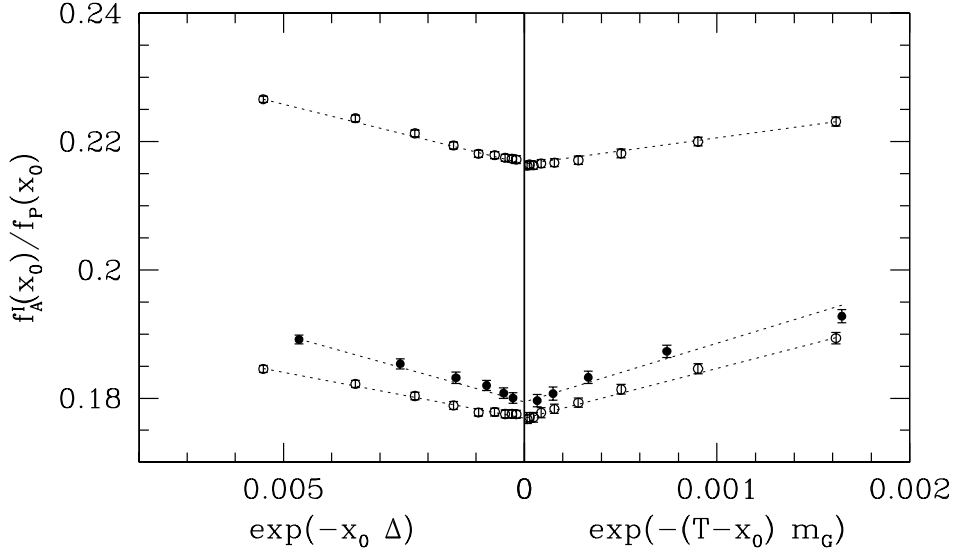,width=14cm}
}
\vspace{-1.2cm}
\caption{\footnotesize
Influence of excited states on the ratio ${\faimpr(x_0) /  \fp(x_0)}$
for $a\approx0.07\,\fm$ (open circles) and
$a\approx0.09\,\fm$  (filled circles). 
\label{f_FGeff}}
\end{figure}
Next, we discuss the bare pion decay constant, $\Fpi^{\rm bare}$,
which is obtained from an average of 
\bes
\Fpi^{\rm bare} = 2(m_{\pi}\, L^3)^{-1/2} \rme^{(x_0-T/2)m_{\pi}} \,
{\faimpr(x_0) \over \sqrt{f_{1}}} \,, \label{e_fpibare} 
\ees 
over the
same interval of $x_0$, with $m_\pi$ taken from the previous
analysis.\footnote{
%
%
The leading correction terms to \eq{e_fpibare} are suppressed by
factors $\mpi/\Delta$ and $\mpi/m_{\rm G}$ compared to
\eq{e_meff_asympt}. However, we did not enlarge the interval in
this case, since $\mpi$ is not very small in
comparison to the other masses.
} 
Values for $\Fpi^{\rm bare}$ are also included in \tab{t_pion}. Note
that in the improved theory which we consider, the renormalized
decay constant is given by $\Fpi=\za(1+\ba \,a\mq)\Fpi^{\rm bare}$
\cite{impr:pap1}.

A further example for the determination of matrix elements is the
combination $\mpi \Fpi^{\rm bare} / \Gpi^{\rm bare}$ which is related
to the ratio $\faimpr/\fp$ via
\bes
 {\faimpr(x_0) \over \fp(x_0)} \approx 
             {m_{\pi}\Fpi^{\rm bare} \over \Gpi^{\rm bare}}
             {      
             \left\{ 1 + \etaa^{\pi}\rme^{-x_0 \Delta } + 
                  \etaa^{0} \rme^{-(T-x_0) m_{\rm G} } 
             \right\}  
             \over 
             \left\{ 1 + \etap^{\pi}\rme^{-x_0 \Delta } + 
                  \etap^{0} \rme^{-(T-x_0) m_{\rm G} } 
             \right\}  
             }\,.
\ees
Again one may average the l.h.s. over a range of $x_0$ where excited
state corrections are negligible. To find the proper range for the
ratio ${\faimpr(x_0) / \fp(x_0)}$, we literally repeat the analysis
performed before for $m_{\rm eff}$. As shown in \fig{f_FGeff} the
corrections of order $\rme^{-x_0 \Delta }$ are of the same order as
before. They originate predominantly from the denominator, since the
PCAC relation predicts
\bes
  \etap^{\pi} = {\Delta + \mpi \over \mpi} \etaa^{\pi}\,. 
\label{e_etaa_etap}
\ees
This enhancement factor compensates for the missing factor ${\Delta
\over \mpi}$ compared to \eq{e_meff_asympt}, and a similar value of
$t_{\rm min}$ has to be chosen (see \tab{t_fitranges}).  Results for
$ m_{\pi} \Fpi^{\rm bare} / \Gpi^{\rm bare}$ are included in
\tab{t_pion}.

\subsection{The vector channel \label{s_vect}}

\begin{figure}[tb]
\hspace{0cm}
\vspace{-5.3cm}

\centerline{
\psfig{file=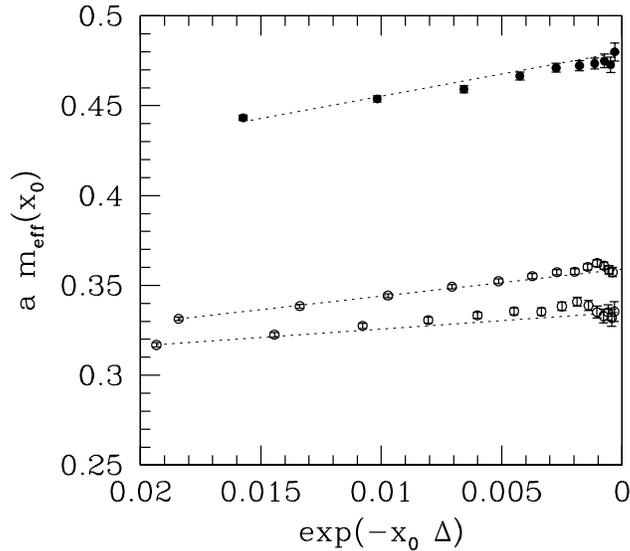,width=14cm}
}
\vspace{-1.2cm}
\caption{\footnotesize
The effective mass of the correlation function $\kvimpr$ against the
leading correction. The meaning of the symbols is as in
\fig{f_FGeff}.
\label{f_m_eff_rho2}}
\end{figure}

In analogy to the pseudoscalar case, the analysis of the correlation
function $\kvimpr$ requires information about the gap in the vector
channel. The effective masses are plotted in \fig{f_m_eff_rho2}.
Here, the estimate for $\Delta$ has been obtained by tuning its
value until a roughly linear behaviour was observed. Thus, unlike
the case of the pseudoscalar, the gap is not known independently
through a separately determined correlation function. 
However, from additional runs performed on a larger volume, we know
that the contribution from excited states has approximately the same
magnitude for $L=2.2\,\fm$, which adds further credibility to the
analysis of the gaps presented here.

Our statistical errors are too large to observe a significant signal
for the glueball contribution at large $x_0$. The value of $t_{\rm
max}$ was therefore determined by requiring that the maximally
allowed glueball amplitude be contained within the statistical
errors in the range of $x_0$ corresponding to  $\exp(-(T-x_0)m_{\rm G})
\leq 0.002$. As before, our results for $m_\rho$ obtained using
the averaging procedure were consistent with single exponential fits
to $\kvimpr$. All parameters and numerical results are listed in the
tables.

\begin{table}[tbh]
\centering

\begin{tabular}{ccr@{.}lr@{.}lr@{.}lr@{.}l}
\hline \\[-1.0ex]
$\beta$ & $\kappa$  & \multicolumn{2}{c}{$a m_\rho$} &
\multicolumn{2}{c}{ref.\cite{impr:qcdsf}} & 
\multicolumn{2}{c}{ref.\cite{impr:roma2_2}} &
\multicolumn{2}{c}{ref.\cite{impr:roma1_spec}}\\[1.0ex] 
\hline \\[-1.0ex]
 6.0 & 0.1338 & 0&508(3) & \blank   & \blank     & \blank    
              \\
    & 0.1342 & 0&480(4) & 0&487(3)   & \blank     & \blank    
              \\[1.0ex]
\hline \\[-1.0ex]
 6.2 & 0.1349 & 0&359(3) & 0&363(4) & \blank     & 0&363(8)
              \\
     & 0.1352 & 0&335(4) & 0&343(5) & 0&353(15)  & 0&335(12)
              \\[1.0ex]
\hline
\end{tabular}
\caption[]{\footnotesize
Vector meson masses 
\label{t_rho}
}
\end{table}

The comparison of our estimates for $m_\rho$ to results employing
standard methods in \tab{t_rho} shows that our numbers are
slightly lower, although the difference is mostly not statistically
significant. In view of the many checks of our analysis, we are
confident that the vector mass has been extracted
correctly. It is well known that in general the determination of the
mass in the vector channel is not easy, so that differences at the
level of around 1.5 standard deviations are not too surprising.

\section{Numerical efficiency \label{s_eff}}

We can now assess the numerical efficiency of our method in relation
to results obtained using conventional techniques. Comparing the
errors in Tables~\ref{t_pion} and~\ref{t_rho}, one has to take into
account that the statistics for the simulation in~\cite{impr:qcdsf}
is approximately the same as ours, whereas in
refs. \cite{impr:roma2_2} and \cite{impr:roma1_spec} the number of
``measurements'' is smaller by roughly a factor~8. If one
compensates for trivial statistical factors, the tables demonstrate
that correlation functions computed in the \SF allow for the
determination of hadron masses with similar precision compared to
conventional methods. This is also the case for ratios of
correlation functions like $\faimpr(x_0)/\fp(x_0)$, which serves to
extract the combination $m_\pi\Fpi^{\rm bare}/\Gpi^{\rm
bare}$~\cite{Allton:1994ps,impr:ukqcd_prep}.

Another relevant issue for the overall precision is the tolerated
maximum contamination by excited states, $\eps$. In order to avoid
the total error to be noticeably affected by systematic effects, the
averaging or fitting intervals must be chosen such that the
statistical error is still significantly larger than $\eps$. It then
turns out that in our approach one can use very small values for
$\eps$ without compromising the statistical accuracy. This is
illustrated by a direct comparison to the results of
ref.~\cite{impr:roma1_spec} in the pseudoscalar channel. From the
formulas in the appendix we have
\bes
   \eps = \chi_{\rm P}^2\,\rme^{\Delta t_{\rm min}},
\ees
for the analysis of \cite{impr:roma1_spec}. Inserting our estimates
for $\Delta$ and $\chi_{\rm P}$ obtained from fits described in the
appendix and the value of $t_{\rm min}$ used in
\cite{impr:roma1_spec}, one obtains $\eps\approx0.6\%$ in the
pseudoscalar channel for ref.~\cite{impr:roma1_spec}. Thus, in our
simulation {\it both\/} the statistical error and the residual
contamination by excited states is smaller by about a factor three.

The overall errors of the observables discussed above are
similar to the ones achievable with standard methods, but perhaps --
as we just argued -- the \SF correlation functions give somewhat
more precise results.  In addition the \SF enables one to compute
the pseudoscalar decay constant with much better precision compared
to what is usually achieved with conventional correlation functions. 
The reason for this is not entirely clear, but it may be because
$\Fpi^{\rm bare}$ defined in eq.~(\ref{e_fpibare}) is obtained from
a straight ratio of correlation functions times a function involving
$m_\pi$ only.

\section{Discussion}

In this paper we have shown how correlation functions with \SF
boundary conditions can be used to compute hadronic quantities like
meson masses and matrix elements with high precision. An integral
part of our analysis was the detailed investigation of the influence
of excited states: in the pseudoscalar channel we have used
independent information about the gap $\Delta$ and the lightest
glueball mass to select the appropriate averaging intervals.

As explained in Section \ref{s_corr}, correlation functions with \SF
boundary conditions decay slowly, resulting in accurate results for
masses and matrix elements. In particular, we have seen that our
method produces {\it very precise} results for the pion decay
constant. One may expect that a similarly good efficiency applies to
pseudoscalar-to-pseudoscalar matrix elements such as $B_{\rm
K}$. Such applications should be investigated in the future.
Furthermore, completely different channels like the nucleon are
accessible and should be tried.

All our detailed investigations have been done for $L\approx
1.5\,\fm$. How does the size of excited state corrections depend on
$L$? For smaller $L$, we expect the dominance by the ground state to
be similar or even better. For significantly larger $L$, however,
our correlation functions might receive bigger contributions from
excited states and the efficiency might deteriorate. We have
investigated also $L \approx 2.2\,\fm$ (keeping $T$ fixed) for two 
different
pairs of $(\beta,\kappa)$. The magnitude of excited 
state corrections is hardly different from our results
on the smaller system. So the location of the window of $x_0$, 
which allows for an extraction of physical masses
and matrix elements, is independent of $L$ between 
$1.5\,\fm \leq L \leq 2.2\,\fm$.  
Even larger values of $L$ can not be reached with
our computing resources but are also not necessary for the
quantities studied here.

Further improvements of the method are possible.  So far we have
used only the simplest implementation of composite boundary fields
in the calculation of $f_1, \faimpr$ and $\kvimpr$. More refined
choices of sources involving the boundary fields, such as tuned
hadron wavefunctions, can surely be made, whilst preserving gauge
invariance at all stages of the calculation. This might further
enhance the efficiency of the method.

\vspace{0.5 cm}

\noindent
{\bf Acknowledgements.} This work is part of the ALPHA collaboration
research programme. We thank DESY for allocating computer time on
the APE/Quadrics computers at DESY-Zeuthen and the staff of the
computer centre at Zeuthen for their support. We are grateful to the
{\it Tor Vergata} group and to the UKQCD collaboration for access to
their data. We also thank R. Horsley and D. Pleiter for a discussion
of their results~\cite{impr:qcdsf} and U. Wolff for a critical reading
of the manuscript.
 
\begin{appendix}

\section{Determination of $\Delta$}

Here we describe the determination of the gap $\Delta$ in the
pseudoscalar channel using correlation functions with periodic
boundary conditions in all four space-time directions. The quark
propagators were made available by the {\it Tor Vergata} group, and
more details about the simulation can be found
in~\cite{impr:roma2_2}. Here we only state that $\Delta$ has been
determined at $\beta=6.2$ on a lattice of size $24^3 \times 48$,
with the $\Oa$ improved action.  In order to distinguish the
correlation function computed using periodic boundary conditions
from those defined within the \SF, we use the letter~`$C$', defining
\bes
C_{\rm PP}(x_0) &=& a^3\sum_{\bf x}\langle P(x)P^\dagger(0)\rangle ,
\label{ppcor} \\
C_{\rm AP}(x_0) &=& 
       a^3\sum_{\bf x}\langle (A_{\rm I})_0(x)P^\dagger(0)\rangle ,
\label{apcor} \\
C_{\rm AA}(x_0) &=& 
  a^3\sum_{\bf x}\langle (A_{\rm I})_0(x)(A_{\rm I}^\dagger)_0(0)
                                \rangle, 
\label{aacor}
\ees
with $P(x)$ and $(A_{\rm I})_0(x)$ as given in
eqs.~(\ref{e_pseudodef}) and~(\ref{e_axialimpdef}).
Furthermore, we here use 
$P^\dagger(x)=-\psibar_{\rm u}(x)\gamma_5\psi_{\rm d}(x)$,
and similarly for $A_{\rm I}^\dagger$. The
spectral decomposition of the correlation functions in terms of the
two lowest intermediate states is
\bes
C_{XX}(x_0) &\approx& \xi_{X}^2\,\rme^{-x_0 m_\pi}
            \times \left\{1+\chi_{X}^2\,\rme^{-x_0\Delta}\right\}
            ,\; X={\rm A,\,P} \\
C_{\rm AP}(x_0) &\approx& \xi_{\rm A}\xi_{\rm P}\,\rme^{-x_0 m_\pi}
            \times \left\{1+\chi_{\rm A}\chi_{\rm P}\,
            \rme^{-x_0\Delta}\right\} ,
\label{expansion}
\ees
when terms proportional to $\exp(-m_{\pi}(T-x_0))$ and $\exp(-m_{G}(T-x_0))$
can be neglected. We now consider the ratio
\bes
R_{\rm AP}(x_0) = {{C_{\rm PP}(x_0) C_{\rm AA}(x_0)}\over
                       {[C_{\rm AP}(x_0)]^2}}, 
\ees
which -- in the same approximation -- is given by 
\bes
R_{\rm AP}(x_0)-1 \approx (\chi_{\rm P}-\chi_{\rm A})^2 \rme^{-\Delta x_0}.
\label{rapapp}
\ees
As in eq.~(\ref{e_etaa_etap}), $\chi_{\rm P}$ and $\chi_{\rm A}$ are
related by the PCAC relation,
\bes
\chi_{\rm P} = {{\Delta+m_\pi}\over{m_\pi}}\chi_{\rm A},
\ees
and hence
\bes
R_{\rm AP}(x_0)-1 \approx \chi_{\rm
P}^2\left({{\Delta}\over{m_\pi}}\right)^2 \rme^{-\Delta x_0}\, .
\label{rapfit}
\ees
%


\begin{figure}[tb]
    \vspace{-2.5cm}
    \psfig{file=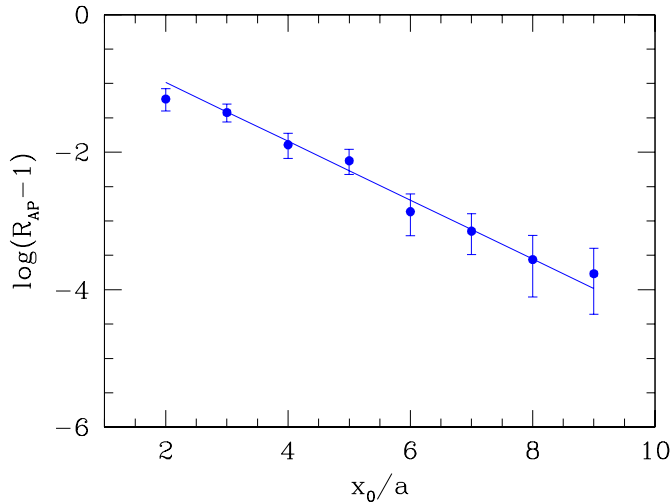,width=10.0cm}
    \vspace{-30pt}
    \caption{\footnotesize{$R_{\rm AP}-1$ for $\kappa = 0.1345$,
    $\beta = 6.2$. The line represents a fit in the window $3a \leq x_0
    \leq 8a$.}} 
    \label{figrap}
\end{figure}

By fitting $(R_{\rm AP}(x_0)-1)$ to the above functional form in the
appropriate range of $x_0$, one can extract the gap $\Delta$. A typical
fit is shown in Fig.~\ref{figrap} from which we obtain the result
quoted in the text,
\bes
   r_0 \Delta  \approx 3.2\,.
\ees
It turns out that $\Delta$ depends very little on the bare quark mass,
so that this result is used in the analysis of correlation functions
in Section~\ref{s_extr} at all values of the quark mass.

\end{appendix}
\bibliography{lattice}        
\bibliographystyle{h-elsevier}   
\end{document}